\documentclass[aps,prd,twocolumn,superscriptaddress,preprintnumbers,floatfix,nofootinbib,10]{revtex4-1}
\usepackage{graphicx}
\usepackage{bm}
\usepackage{times}
\usepackage{slashed}
\usepackage{color}
\usepackage{aas_macros}
\usepackage{slashed}
\usepackage{lipsum}
\usepackage{subfigure}
\usepackage{multirow}
\usepackage{amsmath}
\usepackage{hyperref} 
\usepackage{isotope}
\hypersetup{colorlinks = true, 
citecolor=red,
linkcolor=blue,
filecolor=magenta,      
urlcolor=blue
}
\newcommand{\beq}{\begin{equation}}
\newcommand{\eeq}{\end{equation}}

\newcommand{\ferad}{\rm {\isotope[60]{Fe}}}
\newcommand{\festa}{\rm {\isotope[56]{Fe}}}
\def\lesssim{\buildrel < \over {_{\sim}}}
\def\gtrsim{\buildrel > \over {_{\sim}}}

\begin{document}


\title{Impact of transport modelling on the $^{60}$Fe abundance inside Galactic cosmic ray sources}

\author{Giovanni Morlino}
\email{giovanni.morlino@inaf.it}
\affiliation{INAF/Osservatorio Astrofico di Arcetri, L.go E.~Fermi 5, Firenze, Italy}

\author{Elena Amato}
\email{amato@arcetri.astro.it}
\affiliation{INAF/Osservatorio Astrofico di Arcetri, L.go E.~Fermi 5, Firenze, Italy}
\affiliation{Dipartimento di Fisica e AstronomiaÑUniversit\`a degli Studi di Firenze, Via Sansone, 1, I-50019ÑSesto Fiorentino (FI), Italy}

\date{\today}

\begin{abstract}
The ACE-CRIS collaboration has recently released the measurement of radioactive \ferad\ nuclei abundance in Galactic Cosmic Rays, in the energy range $\sim 195-500$ MeV per nucleon. We model Cosmic Ray propagation and derive from this measurement the \ferad/\festa\ ratio that is expected in the sources of Galactic Cosmic Rays. We describe Cosmic Ray origin and transport within the framework of the {\it disk/halo} diffusion model, namely a scenario in which the matter and the Cosmic Ray sources in our Galaxy are confined to a thin disk, while Cosmic Ray propagation occurs in a much larger halo with negligible matter density. We solve the Cosmic Ray transport equation accounting for spallation reactions, decay and ionization losses as well as advection. We find that the \ferad/\festa\ ratio at the source must be very close to the value detected in the local Cosmic Ray spectrum at Earth, due to the fact that spallation reactions are more effective for \festa\ than for \ferad. Such a result could help identify the sources of Galactic Cosmic Rays.
\end{abstract}

\maketitle

\section{Introduction}		\label{sec:intro}
The question of what the primary sources of Galactic Cosmic Rays (CRs hereafter) are is a very active subject of research. While particle acceleration certainly takes place in Supernova Remnants (SNRs hereafter), there are some important unsettled issues in the paradigm that associates Galactic CRs to Supernova (SN) explosions. Among these is the CR composition, which shows few but relevant peculiarities, likely to hold precious clues both on the main sources and on the acceleration process. Especially important in this sense is the study of nuclear isotopes that are not commonly found in the ISM, like $^{22}$Ne and \ferad. This work focuses on the latter, which is believed to be produced primarily in core-collapse SNe involving stars with mass $\gtrsim 10 M_{\odot}$.

\ferad\ is a radioactive isotope, unstable to $\beta^-$ decay, with a half-life of $2.62 \times 10^6$ years. While SN nucleosynthesis calculations \citep{woosley07,Chieffi-Limongi2013} predict it to be rare, its relatively long half-life has however made it detectable in CRs: a thorough analysis of ACE-CRIS data collected between 1997 and 2014 has revealed the presence of {\ferad\ } nuclei in the energy interval between 195 and 500 MeV/n (energy per nucleon)\citep{Binns+2016}. In this range, the measured {\ferad/\festa\ } ratio is $(4.6 \pm 1.7) \times 10^{-5}$.

In order to derive from this measurement the value of \ferad\ in CR sources, one must first correctly describe two fundamental processes: {\it i}) the particle injection mechanism, that must be able to promote elements from the thermal pool to relativistic energies and {\it ii}) the propagation of CRs from the sources to Earth. 
In the framework of {\it diffusive shock acceleration} theory (DSA), the injection of elements can only depend on the $A/Z$ ratio (where $A$ is the atomic number and $Z$ the effective charge of each specie).  Such a dependence has been invoked to explain the difference between the GCR and solar composition: in particular, the increase of injection efficiency with the ratio $A/Z$ allows one to explain the enhancement of the heavier elements with respect to the lighter ones (among the volatile elements), as well as the mass-independent enhancement of the refractory elements with respect to the volatile ones \citep{Meyer-Drury-Ellison:1997,Ellison-Drury-Meyer:1997}.
Those findings are also in agreement with results from hybrid simulations, where a dependence of injection $\propto (A/Z)^2$ has been found \cite{Caprioli+2017}, implying that the injection of \ferad\ is expected to be enhanced, with respect to that of \festa, by less than $\sim 15\%$.

As far as propagation is concerned, in the \ferad\ discovery paper by the ACE-CRIS collaboration \citep{Binns+2016}, a simplified leaky box model was used to infer a value of $(7.5 \pm 2.9) \times 10^{-5}$ for the \ferad/\festa\ ratio in the CR sources. Such a high value would clearly imply that the \ferad\ observed in CRs cannot originate from the acceleration of the average interstellar medium (ISM), where the relative abundance of \ferad\ is much lower. In fact, the \ferad\ abundance has been measured in the ISM through the detection of $\gamma$-ray lines produced by its decay. The best available measurement comes from the spectrometer on board the INTEGRAL mission \citep{Wang+2007} and returns a \ferad/\festa\ ratio of $\sim 3 \times 10^{-7}$ \cite[see also][]{Trappitsch-Boehnke+2018}\footnote{It is interesting to note that the estimated value of $^{60}$Fe$/^{56}$Fe ratio at the time of formation of the Solar System is even lower, being $\sim (3.8 \pm 6.0) \times 10^{-8}$ \cite{Trappitsch-Boehnke+2018}.}.

In a time when several different aspects of the standard scenario for the origin and propagation of CRs are being questioned, pushed by both new data and theoretical developments \cite[see][for a review]{Amato-Blasi:2017,Aloisio-Blasi+2018}, gamma-ray observations \citep{Aharonian+2018} have recently revived the suggestion by \cite{Cesarsky-Montmerle1983} that the winds of massive stars might be important (if not the primary) CR sources. Such a scenario would imply a paradigm shift, but what is interesting in the context of this work is that one of the possible tests consists exactly in the CR abundance of \ferad: stellar wind material has essentially the same composition as the galactic average, and hence a large \ferad\ abundance in CR sources disfavours these winds as the main CR contributors, at least at low energies (below 1 GeV/n), where these measurements are available.

However, before deriving any firm conclusion about the \ferad/\festa\ ratio at the sources, it is important to make sure that CR  propagation is correctly accounted for. In fact, the leaky-box model adopted by \citep{Binns+2016} in the discovery paper is not appropriate to describe the propagation of unstable nuclei whose decay time is smaller than the escape time from the Galaxy \cite[see, e.g.][]{Ptuskin-Soutoul1998, Ptuskin-Soutoul1998Rev}. To overcome this difficulty and provide a more reliable estimate of the \ferad\ abundance in CR sources, here we model the Iron propagation using the {\it disk/halo model}  \citep{Jones+2001}, where the Galaxy geometry is taken into account in a more realistic way. Such a model has been successfully applied to explain the spectrum of several CR species \cite[see, e.g.,][]{Jones+2001} also in the context of self-generated turbulence \citep{Aloisio-Blasi2013, Aloisio-B-S:2015,Evoli-Blasi+2018}. In general, for stable nuclei, the disk is treated as infinitely thin: this approximation allows one to derive an analytical solution by means of the weighted slab technique, which, compared to numerical techniques often used to solve the CR transport equation \citep{DRAGON:2017,GALPROP:1998}, has the advantage of providing a more immediate picture of the underlying physics. However, the thin disk approximation becomes, in principle, inappropriate for unstable nuclei, when the propagation length-scale becomes of the order of the disk thickness. For this reason here we also check its results against the solution obtained for a finite disk size, quantifying the difference between the two approaches.

The paper is organised as  follows. In \S~\ref{sec:thin_disk} and \ref{sec:thick_disk} we present the solution for the CR spectrum of stable and unstable nuclei for a thin and a thick Galactic disk, respectively. In \S~\ref{sec:grammage} we introduce the grammage and in \S~\ref{sec:transport} we discuss the two different transport models we assume for our calculations. In \S~\ref{sec:results} we present quantitative results for the $^{60}$Fe$/^{56}$Fe ratio in CR sources, in both propagation scenarios we consider. Finally, we discuss the differences between our approach and the leaky-box model in \S~\ref{sec:leaky-box} and our conclusions in \S~\ref{sec:conc}.

\section{The CR distribution function} \label{sec:model}
In this section we solve the CR transport equation within a disk/halo model of the Galaxy, considering both a thin (\S~\ref{sec:thin_disk}) and a thick (\S~\ref{sec:thick_disk}) disk. While the latter allows a more appropriate treatment of the case of unstable nuclei, the former, being simpler, serves the purpose of illustrating the role of the different physical processes determining the CR spectra. In addition, it also allows one to introduce and directly quantify the accumulated grammage, as we discuss in \S~\ref{sec:grammage} and \ref{sec:results}.

\subsection{The thin disk solution} 	\label{sec:thin_disk}
The transport equation for Iron nuclei that undergo spallation, decay and also ionization losses is written as: 
\begin{eqnarray} \label{eq:transport0}
  -\frac{\partial}{\partial z} \left[ D \frac{\partial f}{\partial z} \right] 
  + u \frac{\partial f}{\partial z}
  - \frac{d u}{dz} \frac{p}{3} \frac{\partial f}{\partial p} 
  + \frac{f}{\tau_{\rm sp}} +  \frac{f}{\tau_d} \\ \nonumber 
  + \frac{1}{p^2}  \frac{\partial}{\partial p} \left[  \dot{p}_{\rm ion} \, p^2 f \right] 
  = q(p,z)  \,,
\end{eqnarray}
where $z$ is the height above or below the disc, located at $z=0$; $D(p,z)$ is the diffusion coefficient; $u(z)$ is the advection velocity, directed along $z$; and $q$ is the injection rate CR sources provide per unit volume. Finally, $\dot p_{\rm ion}$ describes ionization losses, while $\tau_{\rm sp}$ and $\tau_d$ are the spallation and decay timescales, respectively.  Notice that we are not including a source term coming from the spallation of heavier elements, because for Iron this is completely negligible. Moreover we are neglecting the diffusion in momentum space because second order acceleration is found, {\it a posteriori}, to be irrelevant, in the propagation model we consider (see the end of \S~\ref{sec:transport}).

We simplify Eq.~\eqref{eq:transport0} adopting a 1D slab model as described in \cite{Jones+2001}: the CR sources are located only inside a thin disk of half-thickness $h$, while the confining volume is a thicker halo, with half-thickness $H \gg h$. The majority of matter is concentrated inside the disk, where the gas density is $n_d$. The gas density in the halo, $n_h$, is assumed to be negligible, so that spallation and ionization losses only occur inside the thin disk\footnote{Notice that this assumption is violated when $n_h H \gtrsim n_d h$. When this happens, spallation in the halo cannot be neglected anymore.}. 
We further assume that $D$ and $u$ are constant in the halo with $u=u_0 (2\Theta(z) - 1)$.
In such a simplified model, the 1D transport equation reduces to:
\begin{eqnarray} \label{eq:transport1}
  - \frac{\partial}{\partial z} \left[ D \frac{\partial f}{\partial z} \right] 
  + u_0 \frac{\partial f}{\partial z}
  - \frac{2}{3} u_0 p \frac{\partial f}{\partial p} \delta(z)
  + \frac{2 h \delta(z)}{\tau_{\rm sp}} f   +  \frac{f}{\tau_d} 
    \nonumber  \\
  - \frac{1}{p^2}  \frac{\partial}{\partial p} \left[  \frac{p^3}{\tau_{\rm ion}} \, f \right] 2 h \delta(z)
  = 2 h q_0(p) \delta(z) \,, \hspace{0.5cm}
\end{eqnarray}
where we introduced $\tau_{\rm ion} = -p/\dot{p}_{\rm ion}$. Notice that often the collisional terms are written as a function of the disk column density which is a measured quantity, $\mu= 2 h n_d m = 2.4 \,{\rm mg \, cm^{-2}}$ \citep{ferriere98}. Hence, for both spallation and ionization losses, we can write $2 h/\tau = \mu v \sigma/m$, with $m=1.4m_p$, the average mass of gas particles.
The spallation and ionization cross sections we use are reported in Appendix \S~\ref{sec:losses}.

In order to solve Eq.~\eqref{eq:transport1}, we proceed using a standard technique: we first solve the equation for $z>0$, where injection, spallation and ionization processes are absent; then we look for the solution at $z=0$, integrating Eq.~\eqref{eq:transport1} around the disk discontinuity. Above and below the disc, the transport equation reads:
\begin{equation} \label{eq:halo}
  D(p)\frac{\partial^2 f}{\partial z^2}
  - u_0 \frac{\partial f}{\partial z}  
  -  \frac{f}{\tau_d} = 0 \,,
\end{equation}
which is a linear second order differential equation whose general solution is 
\begin{equation}
  f= A e^{\alpha_+ z} + B e^{\alpha_- z} \ ,
  \label{eq:sol1}
\end{equation}
where $\alpha_{\pm}$ are the solutions of the second order algebraic equation $D \alpha_\pm^2 - u_0 \alpha_\pm -1/\tau_d=0$. The coefficients $\alpha_\pm$ are then:
\begin{equation}
  \alpha_{\pm} = \frac{u_0}{2 D} \left[ 1 \pm \sqrt{1+ \frac{4 D}{u_0^2 \tau_d} }\right] 
  \equiv \frac{u_0}{2 D} \left[ 1 \pm \Delta \right]\,,
  \label{eq:alpha}
\end{equation}
where we have introduced the dimentionless quantity $\Delta$ that can also be written as a function of the time scales involved in the propagation process, namely
\begin{equation}
  \Delta = \sqrt{1+ 4 \tau_{\rm adv}^2/ \left(\tau_{\rm diff} \tau_d \right) } \,
\end{equation}
with $\tau_{\rm diff} = H^2/D$ and $\tau_{\rm adv}= H/u_0$. Clearly $\Delta\rightarrow 1$ for $\tau_d \gg \tau_{\rm adv} , \, \tau_{\rm diff}$.
Now, the constants $A$ and $B$ in Eq.~\eqref{eq:sol1} are determined by imposing the boundary conditions at the Galactic disk and at the edge of the halo: $f(p, z=0) = f_0(p)$ and $f(p, z=\pm H) = 0$. The final solution, for $z>0$, reads:
\begin{equation}
  f(z,p) = f_0(p) \frac{1-e^{u_0 \Delta (z-H) / D}}{1-e^{-u_0\Delta H/D}} e^{u_0(1-\Delta) z/2D}\, ,
\label{eq:f_halo}
\end{equation}
which in the case of stable elements ($\Delta=1)$ reduces to the well know solution
\begin{equation}
  f_{\rm stable}(z,p) = f_0(p) \frac{1 - e^{u_0 (z-H)/D} }{1 - e^{-u_0 H/D}} \,.
\end{equation}
The distribution function inside the disc, $f_0(p)$, can be obtained by integrating Eq.\eqref{eq:transport1} between $0^-$ and $0^+$:  
\begin{eqnarray} 	\label{eq:f0_1}
  - 2 \left[ D \frac{\partial f}{\partial z} \right]_{z=0^+} 
  - \frac{2u_0}{3} p \frac{\partial f_0(p)}{\partial p} 
  + \frac{2h}{\tau_{\rm sp}} f_0(p) 				\nonumber \\
  - \frac{2 h}{p^2} \frac{\partial}{\partial p} \left[  \frac{p^3}{\tau_{\rm ion}} f_0 \right] 
  = Q_0\, , 
\end{eqnarray}
with $Q_0(p)=2hq_0(p)$. 
The quantity $D \partial f/\partial z |_{0^+}$ represents the diffusive flux at the disk position and can be obtained deriving Eq.~\eqref{eq:f_halo} with respect to $z$, namely:
\begin{eqnarray} \label{eq:Ddfdx}
  \left[ D \frac{\partial f}{\partial z} \right]_{z=0} 
  &=& f_0 \frac{u_0}{2} \frac{(1 - \Delta) - (1 + \Delta) e^{-u_0 \Delta H/D} }{1-e^{-u_0 \Delta H/D}} \\ \nonumber
  &\equiv& - f_0(p) \frac{u_0}{2}  \xi(p) \,.
\end{eqnarray} 
The quantity $\xi(p)$ is a measure of the gradient of the distribution function in units of $D/u_0$. Its meaning is easily appreciated in a few limiting cases. Let us introduce the scale-length $L$ such that:
\begin{equation} \label{eq:Ddfdx2}
  \left[ D \frac{\partial f}{\partial z} \right]_{z=0} \simeq -f_0 \frac{D}{L} \,.
\end{equation} 
In the case of stable nuclei, $\Delta=1$ and $\xi= 2/(e^{u_0H/D} -1)$. If we now consider the diffusion dominated case, i.e. $D\gg u_0H$, we find $\xi\rightarrow 2 D/(u_0 H)$, which shows that the gradient of the distribution function is on a scale $L=H$. On the other hand, in the advection dominated case $\xi\rightarrow 0$, and the scale-length is $L\to \infty$. Finally, in the case of unstable elements, if $\tau_d\ll 4D/u_0^2$ and $\tau_d\ll t_{\rm diff}$, are both satisfied, one finds $\xi\to \Delta=\sqrt{4D/(u_0^2\tau_d})$ and $L=\sqrt{D \tau_d}$.

Using Eq.~\eqref{eq:Ddfdx}, we can recast Eq.~\eqref{eq:f0_1} as follows
\begin{equation} \label{eq:f0_2}
  p \frac{\partial f_0(p)}{\partial p} = 
  \frac{ \lambda_1(p) f_0(p) - Q_0(p)}{\lambda_2(p)}
\end{equation} 
where
\begin{eqnarray} 
  \lambda_1(p) &\equiv&  \xi(p) u_0 + \frac{2 h}{\tau_{c1}}   \,,		\label{eq:lambda1} \\
  \lambda_2(p) &\equiv&  \frac{2}{3} u_0 + \frac{2 h}{\tau_{\rm ion}}	\label{eq:lambda2} 
\end{eqnarray} 
and 
\begin{eqnarray}
\tau_{c1}^{-1}     &=&\tau_{\rm sp}^{-1}+(\alpha_{\rm ion}-3)\tau_{\rm ion}^{-1} \,,  \\
\alpha_{\rm ion} &=& \frac{d \ln(\tau_{\rm ion})}{d \ln(p)}\ .
\label{eq:tc1}
\end{eqnarray}

Eq.~\eqref{eq:f0_2} is a first order differential equation in $p$ whose solution can be found as:
\begin{equation} \label{eq:f0_sol}
  f_0(p) =  \int_p^{\infty}  \frac{dp'}{p'} \frac{Q_0(p')}{\lambda_2(p')} \,
  	        \exp \left[ - \int_p^{p'}  \frac{\lambda_1(p'')}{\lambda_2(p'')} \frac{dp''}{p''} \right]\, .
\end{equation} 
Eq.~\eqref{eq:f0_sol} shows that $f_0(p)$ is formed by particles injected with momentum $p' \geq p$, that lose energy down to $p$ during propagation.
The energy decrease is due to both adiabatic and ionization losses: particles loose energy each time they cross the disk because of adiabatic expansion and ionizing collisions. Clearly energy losses are important only at low energies, i.e. when $D \lesssim L/u_0$, which generally occurs for $E\lesssim few$ GeV for standard propagation parameters. In the opposite limit, for energies such that $D \gg L/u_0$, we have $\lambda_1/\lambda_2 \gg 1$ and the exponential function reduces to a Dirac-$\delta$:
\begin{equation}  
  \exp \left[ -\int_p^{p'} \frac{\lambda_1(p'')}{\lambda_2(p'')} \frac{dp''}{p''} \right] 
  \rightarrow \frac{p' \lambda_2(p')}{\lambda_1(p')} \delta(p-p') \,.
\end{equation} 
In this limit Eq.~\eqref{eq:f0_1} reduces to $f_0(p) \approx  Q_0(p)/\lambda_1(p)$, which reproduces the standard result for stable nuclei when spallation and ionization are neglected: this is simply $f_0(p) =Q_0 H/(2D)$.

\subsection{The thick disk solution} 	\label{sec:thick_disk}
As mentioned above, unstable elements whose propagation length, $L_{\rm diff}=\sqrt{D \tau_{d}}$, is of the order of, or smaller than, the disk size, are not accurately described by the infinitely thin disk solution. For \ferad,   the diffusion length is $\sqrt{D\tau_d} \simeq 140 \,D_{27}^{1/2}$ pc at $E \simeq 500$ MeV/n, where $D_{27}$ is the diffusion coefficient in units of $10^{27}$ cm$^2$ s$^{-1}$. Therefore, in the energy range of CRIS measurements, the diffusion length of \ferad\ is comparable with the disk size.

In order to compare our model results with CRIS data, we are then forced to take into account the finite size of the disk. The solution of this problem has long been known for cases when ionization losses can be neglected \cite[see, e.g.][]{Ginzburg-Ptuskin1976,Berezinskii+1990}. However ionization plays an important role in the energy range we are interested in, so in the following we present our own solution of Eq.~\eqref{eq:transport0}, for the case of a thick disk (of half thickness $h$) with a uniform distribution of gas and CR sources.

The steps towards solving Eq.~\eqref{eq:transport0} are similar to the ones in the previous section. We first find a solution for the halo ($h<z<H$), where losses are absent and only the decay term is important. Then we obtain a solution valid inside the disk ($|z|<h$), including spallation and ionization. Finally, we find the CR spectrum at the Earth location ($z=0$), by integrating Eq.~\eqref{eq:transport0} between $0^-$ and $0^+$. The transport equation in the halo is the same as in Eq.~\eqref{eq:halo},  hence the solution for $h < z < h+H$ is the same as Eq.~\eqref{eq:f_halo}, except that $z$ has to be replaced with $z-h$ and $f_0$ with $f_h=f(h,p)$. We then find the solution for the distribution function in the halo as:
\begin{equation}
f_{\rm out}(z,p) = f_h(p) \frac{1-e^{u_0 \Delta (z-h-H) / D}}{1-e^{-u_0\Delta H/D}} e^{u_0(1-\Delta) (z-h)/2D}.
\label{eq:f7b}
\end{equation}

Concerning the solution inside the disk, we need to include the spallation and ionization terms. The latter term is, in principle, more delicate to handle because it contains the momentum derivative of $f$. We write this term as:
\begin{eqnarray}
  \frac{1}{p^2} \frac{\partial}{\partial p} \left[ p^2 \dot{p_{\rm ion}} f \right]
  &=& \frac{f}{\tau_{\rm ion}} \left[ \frac{\partial \ln \tau_{\rm ion}}{\partial \ln p} 
 		- \frac{\partial \ln f}{\partial \ln p} - 3 \right]		\nonumber \\
  &=& \frac{f}{\tau_{\rm ion}} \left[ \alpha_{\rm ion} + \alpha - 3 \right]	  \,.	
\end{eqnarray}
The latter expression is linear in $f$ apart from the spectral slope $\alpha$ which, however, can be approximated as constant for the purposes of the present work. We can then define a compound timescale (analogous to that in Eq.~\eqref{eq:tc1}), 
\begin{equation} \label{eq:tau_c}
  \tau_c^{-1} \equiv \tau_d^{-1} + \tau_{\rm sp}^{-1} + \left( \alpha_{\rm ion} + \alpha - 3 \right) \tau_{\rm ion}^{-1}\ ,
\end{equation}  
such that the transport equation in the disk can be rewritten as
\begin{equation} \label{eq:disk}
  - D(p)\frac{\partial^2 f_{\rm in}}{\partial z^2}
  + u_0 \frac{\partial f_{\rm in}}{\partial z}  
  + \frac{f_{\rm in}}{\tau_c} = q(p) \,.
\end{equation}
The latter equation is completely analogous to Eq.~\eqref{eq:halo}. The two coefficients that appear in its solution, together with $f_h(p)$ appearing in Eq.~\eqref{eq:f7b}, can all be determined imposing the following boundary conditions: $f_{\rm in}(0,p) = f_0(p)$, $f_{\rm in}(h,p)=f_{\rm out}(h,p)$ and $ \partial_z f_{\rm in}(h,p) = \partial_z f_{\rm out}(h,p)$, where the last two conditions entail the continuity of the particle distribution function and its flux at the boundary between the disk and the halo, under the assumption that the diffusion coefficient is the same in the two regions.

The distribution function at the centre of the disk, $f_0(p)$, is obtained again integrating Eq.~\eqref{eq:transport0} between $0^-$ and $0^+$, which gives
\beq 	\label{eq:f0_3}
  - 2 \left[ D \frac{\partial f_{\rm in}}{\partial z} \right]_{z=0^+} 
  - \frac{2u_0}{3} p \frac{\partial f_0}{\partial p} 
  = 0\ , 
\eeq
where the term in the square brackets can be obtained deriving the solution of Eq.~\eqref{eq:disk} with respect to $z$.
The final differential equation for $f_0$ has the same form as Eq.~\eqref{eq:f0_2}, namely 
$p \partial_p f_0(p) = \Omega_1 f_0 - \Omega_2$, and the solution reads
\begin{equation} \label{eq:f0_sol_thick}
  f_0(p) =  \int_p^{\infty}  \frac{dp'}{p'} \Omega_2(p') \,
  	        \exp \left[ - \int_p^{p'}  \Omega_1(p'') \frac{dp''}{p''} \right]\, 
\end{equation} 
where
\begin{eqnarray}
  \Omega_1 = \frac{3}{2} \sum_{+,-}  \frac{-1 \mp \Delta_{\rm in}}{1- \frac{1 \pm \Delta_{\rm in} + \xi}{1 \mp \Delta_{\rm in} + \xi}
  		      						\, e^{ \pm (\alpha_{+} + \alpha_{-})h }}  \hspace{2.4cm} \,,  \\
  \Omega_2 = \frac{3 D q}{u_0^2 \Delta_{\rm in}} \,
  		           \bigg\{ \frac{2}{3} \Omega_1
		      		\left[ \frac{e^{-\alpha_{+} h} - 1}{1 + \Delta_{\rm out}} - \frac{e^{-\alpha_{-} h} - 1}{1 - \Delta_{\rm out}} \right] +
			   	\nonumber \hspace{0.7cm} \\
				 + \, e^{-\alpha_{+} h} - e^{-\alpha_{-} h}
		      	  \bigg\}  \,, \hspace{2.6cm}  
\end{eqnarray}			  
and
\begin{eqnarray}
  \xi = \frac{\Delta_{\rm out} - 1 + (1+ \Delta_{\rm out}) e^{-\frac{u_0 \Delta_{\rm out} H}{D}} }
  		{1 - e^{-\frac{u_0 \Delta_{\rm out} H}{D}}}     \,, 		      \hspace{2.5cm} \\
  \alpha_{\pm}   = \frac{u_0}{2 D} \left[ 1 \pm \Delta_{\rm in} \right]	   \,, \hspace{4.55cm}\\		
  \Delta_{\rm in} = \sqrt{1 + \frac{4 D}{u_0 \tau_c}}  \,, 		               \hspace{0.5cm}
  \Delta_{\rm out} = \sqrt{1 + \frac{4 D}{u_0 \tau_d}}  \,.			       \hspace{1.5cm}
\end{eqnarray}

The integral in Eq.~\eqref{eq:f0_sol_thick} is performed using the numerical technique presented in \cite{Bresci+2019}. We verified that the solution \eqref{eq:f0_sol_thick} gives the same result as Eq.~\eqref{eq:f0_sol} when $h/H \rightarrow 0$.

\section{Grammage} \label{sec:grammage}
While the thick disk solution provides a more accurate description of CR propagation in the situation we are considering, the thin disk approximation is more useful if one wants to discuss propagation in terms of the grammage that particles accumulate.  Following \cite{Jones+2001,Aloisio-Blasi2013}, the grammage can be derived  rewriting Eq.~\eqref{eq:f0_1} in terms of $I(E)$, namely the particle flux as a function of kinetic energy per nucleon $E$. The equality $I(E) dE = v p^2 f_{0}(p)dp$ implies that $I(E) = A p^2 f_0(p)$, $A$ being the atomic mass number of the nucleus.
With such a substitution, Eq.~\eqref{eq:f0_1} can be rewritten as
\begin{eqnarray}\label{eq:IE1}
(\xi+2)u_0I(E)-
\frac{v}{Ac}\frac{d}{dE}\left\{
\left[p\left(\frac{2}{3} u_0+\frac{2h}{\tau_{\rm ion}}\right)\right]I(E)\right\}\nonumber \\
	+ \frac{2h}{\tau_{\rm sp}} I(E)=2hq_0 A p^2\, \hspace{1cm}
\end{eqnarray}
where we have used the definition of $\xi$ from Eq.~\eqref{eq:Ddfdx} and $p=\sqrt{E(E+2m_pc^2)}/c$.
Rather than solving Eq.~\eqref{eq:IE1} explicitly, we want to focus on the differences between stable and unstable nuclei in terms of accumulated grammage, which provides immediate insight on how the different isotopes are affected by propagation.

To this purpose, we recast Eq.~\eqref{eq:IE1} in a more useful form by introducing the disk column density $\mu$ mentioned in \S~\ref{sec:thin_disk} and write:
\begin{eqnarray} \label{eq:I(E)}
  \frac{I(E)}{X(E)} + \frac{d}{dE} 
  	\left\{ 
  		\left[ 
			\left( \frac{dE}{dx} \right)_{\rm ad}  +  \left( \frac{dE}{dx} \right)_{\rm ion}
		\right] \, I(E)
	\right\}   \nonumber  \\
	+  \frac{\sigma_{\rm sp} I}{m_p}
	= Q(E)  \,, \hspace{1cm}
\end{eqnarray}
where
\begin{equation} \label{eq:X(E)}
  X(E) = \frac{\mu v}{u_0} \, \frac{1-e^{-u_0\Delta H/D}}{(1+\Delta) - (1-\Delta) e^{-u_0\Delta H/D}}
\end{equation} 
is the grammage for nuclei with kinetic energy per nucleon $E$,
\begin{equation} \label{eq:dEdt_ad}
  \left( \frac{dE}{dx} \right)_{\rm ad} = - \frac{2 u_0}{3 \mu c} \, \sqrt{E (E+2m_p c^2)}
\end{equation} 
is the rate of adiabatic energy losses,
\begin{equation} \label{eq:dEdt_ion}
  \left( \frac{dE}{dx} \right)_{\rm ion} = - \frac{2hp}{\mu \tau_{\rm ion}}=
  -\frac{\sigma_{\rm ion} v \sqrt{E (E+2m_p c^2)}}{cm_p}
\end{equation} 
is the rate of energy losses due to ionization, and 
\begin{equation} \label{eq:Q(E)}
   Q(E) = \frac{2 h}{\mu v} \, A p^2 q_{0}(p)
\end{equation} 
is the source term.

Notice that with respect to the result presented by \cite{Jones+2001,Aloisio-Blasi2013}, here the grammage $X(E)$ has a more complicated expression because it also accounts for decay. However, Eq.~\eqref{eq:X(E)} immediately shows the asymptotic behaviour of the grammage in three different cases: advection-dominated, diffusion-dominated and decay-dominated regimes. The corresponding approximate expressions are:
\begin{eqnarray}
  &X& = \frac{\mu v}{2 u_0}  \hspace{1.75cm}  {\rm when} \; \tau_{\rm adv} \ll \tau_{\rm diff}\,, \tau_{d} \,; \label{eq:xadv} \\
  &X& = \frac{\mu v H}{2 D}  \hspace{1.58cm}  {\rm when} \; \tau_{\rm diff} \ll \tau_{\rm adv}\,, \tau_{d} \,;  \label{eq:xdiff} \\
  &X& = \frac{\mu v \tau_d}{2\sqrt{ D \tau_d} }
  					  \hspace{1.2cm}  {\rm when} \; \tau_{d} \ll \tau_{\rm diff}\,, \tau_{\rm adv} \,.\label{eq:xdec}
\end{eqnarray} 
These expressions will be useful for the interpretation of the ratio between the isotopes that we discuss in \S~\ref{sec:results}.

\section{Transport model} \label{sec:transport}
\begin{table}[t!!!!]
\caption{Values of the parameters used in the model.}
\begin{center}
\begin{tabular}{c|ccc|ccccc|c}
  $u_0$	 &     $h$	&   $H$   &  $n_{p,\rm disk}$    &	  $D_0$ 	             & $\delta$ &  $\Delta\delta$ & $s$ & $R_{b}$  & $\gamma$	\\ 
  (km/s)	 &     (pc)  	&   (kpc)  &  (cm$^{-3}$)           &  ($\rm cm^{2} s^{-1}$)  &    	     & 			       & 	 &   (MV)	  &		         \\
  \hline \hline
  5.0		&	150	&	7     &       1.5			&	$3.08 \times 10^{28}$ &    0.54   &  	 0.2	       &  0.1  &  312       &      4.3 		\\
  \hline
\end{tabular}
\end{center}
\label{table1}
\end{table}
We now go back to the CR transport equation for a thick disk and discuss its full solution, Eq.~\eqref{eq:f0_sol_thick}. Since experimental results are usually presented in terms of the particle flux as a function of kinetic energy per nucleon, $I(E)$, we will present our result in this form.

The model has several parameters that need to be fixed in order to provide a meaningful estimate for the {\ferad/\festa\ } ratio at the source. The parameters to be fixed are $H$, $h$, $u_0$, the diffusion coefficient $D$ and the injection spectrum $q_0(p)$. Their combination can be constrained by fitting the observed spectra of primary and secondary nuclei, and their ratios. As fiducial values we decided to use those estimated by \cite{Evoli+2019} and \cite{Evoli[Be]+2020}, where the authors used a 1D model in the thin disk approximation identical to the one described in \S~\ref{sec:thin_disk}. Table~\ref{table1} summarises the best fit values of the model free parameters. We discuss them below.

In order to account for the spectral break observed in all CR spectra at a rigidity $\sim 300$ GV, in \cite{Evoli+2019} the diffusion coefficient is described by the following functional form
\begin{equation} \label{eq:D(R)}
  D(R) = 2 u_0 H + \beta D_0 \, 
  		\frac{\left( R/{\rm GV}\right)^{\delta}}{[ 1 + ( R/R_b )^{\Delta\delta/s} ]^s} \, ,
\end{equation} 
where $\beta=v/c$ with $v$ the velocity of the particle of rigidity $R$, $D_0$ is the value of the diffusion coefficient at $R= 1$ GV and the break is described in terms of $s$, $\Delta\delta$ and $R_b$, which are, respectively, a smoothing parameter, the magnitude  and  the characteristic rigidity of the break. Even if Eq.~\eqref{eq:D(R)} reflects a phenomenological approach, its form has been inspired by previous works \cite{Blasi-Amato-Serpico:2012, Aloisio-B-S:2015, Aloisio-Blasi2013}, where the diffusion is described using two different sources of scattering: the externally generated turbulence, which dominates the transport at high rigidities, and the CR self-generated turbulence, which dominates, instead, at lower rigidities. In other words, the spectral break of $D$ reflects the transition between these two regimes. 
At small rigidities, namely for $R \lesssim 1$ GV, where advection becomes important, the diffusion flattens to $D \rightarrow 2 u_0 H$. The presence of such a plateau is also a consequence of the self-generated turbulence as found in Ref.~\cite{Recchia+2016} and reflects the fact that advection and diffusion are equally important (or, in other words, pure advection never dominates). It is worth noticing that in pure diffusion models (where advection and reacceleration are neglected) the existence of such a plateau is in any case required by the data \citep{Jones+2001}.

In \cite{Evoli[Be]+2020} all parameters are fixed by performing a global fit over the AMS-02 data, in particular using the spectrum of $p$, He, C, N, O  plus the ratios Be/C, B/C, Be/O and B/O. Notice that $D_0$ and $H$ cannot be determined separately based on the flux of stable secondary and primary CRs alone, because this only constrains the ratio $D_0/H$. In order to disentangle the two quantities, it is necessary to use unstable elements. Unfortunately, measurements of unstable isotopes are available only at very low energies and AMS-02 is not able to distinguish between isotopes of the same element. However, the CR Beryllium is composed by a non negligible fraction of $^{10}$Be, whose half-life is $1.51$ Myr, so that the decay signature is clearly visible in the total Be flux. \cite{Evoli[Be]+2020} used the Beryllium flux measured by AMS-02 to fix the halo thickness, providing a best fit of $H = 7$ kpc. In our analysis we will adopt such a value, which is, however, slightly larger than the one usually adopted in the literature (closer to $\sim 4-5$ kpc \citep{Jones+2001}). In the next section we will comment on how our results are affected by the halo thickness. The remaining parameter values are the ones reported in Table~\ref{table1}.

\begin{figure}
\begin{center}
\includegraphics[width=0.48\textwidth]{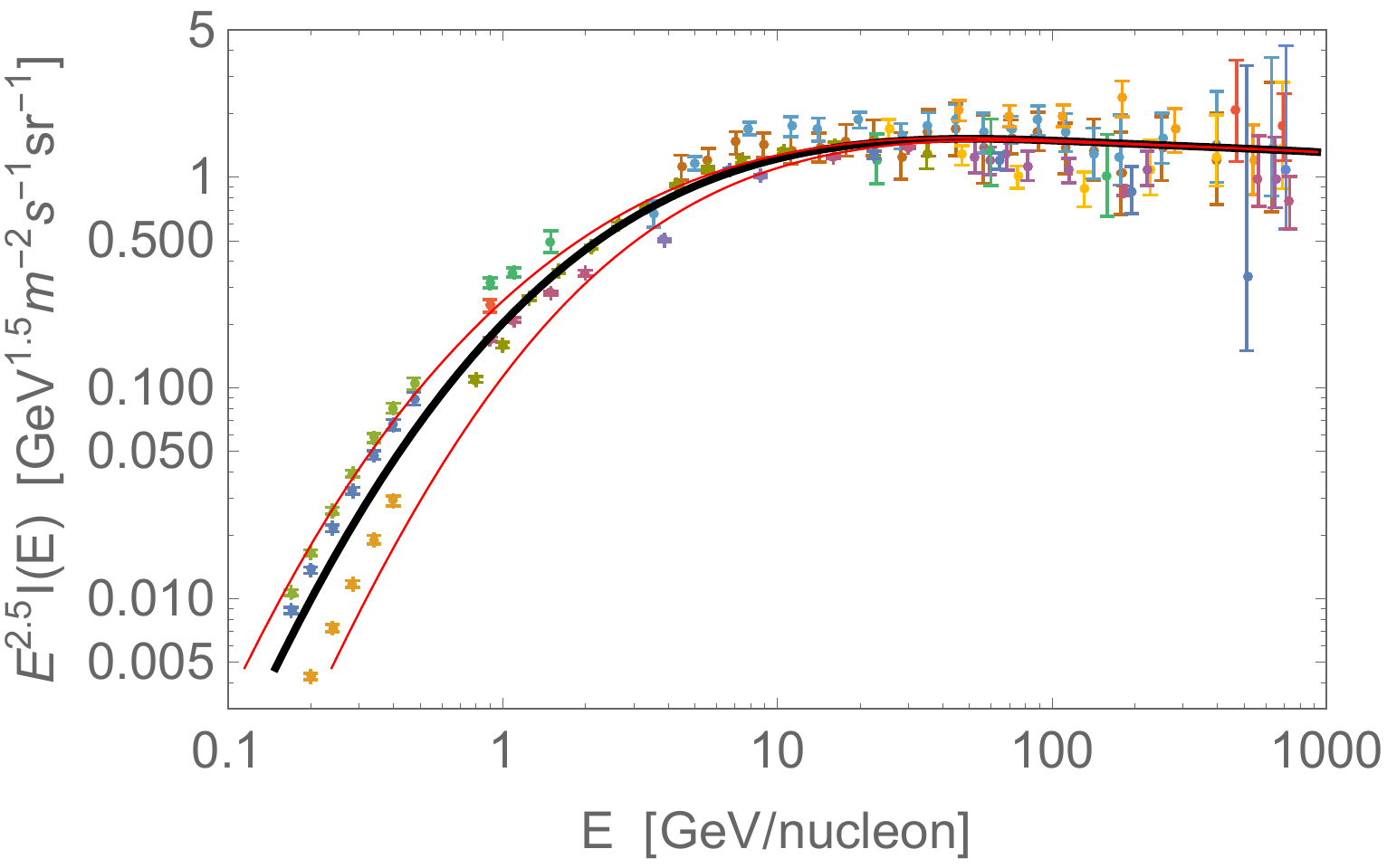}
\end{center}
\caption{Model predicted spectrum of Iron compared with data from different experiments. The solid-black line is calculated using a solar modulation potential equal to $\Phi= 453$ MV, as estimated by \cite{Binns+2016} while the upper and lower thin-red lines have $\Phi= 250$ and 1000 MV, respectively. Data are taken from the Cosmic Ray Database \cite[][\texttt{http://lpsc.in2p3.fr/crdb/}]{Maurin2014} and include all experiments performed after 1980.}
\label{fig:Fe_spectrum}
\end{figure}

The CR spectrum injected by the sources is assumed to be a simple power law $q_{0,i}(p)\propto p^{-\gamma_i}$ where the spectral index $\gamma_i$ can differ for different species. The best fit gives $\gamma_{p}=4.35$, $\gamma_{\rm He}= 4.25 $ while all heavier elements have the same slope $\gamma_{\rm CNO} = 4.3$. In \cite{Evoli[Be]+2020} the Fe spectrum is not taken into account, because no such data have been released by the AMS-02 collaboration so far, hence we adopt here $\gamma_{\rm Fe}= 4.3$, as for CNO elements.
A comparison between the predicted spectrum  and existing data is shown in Figure~\ref{fig:Fe_spectrum}. Considering that the error bars above $\sim 10$ GeV/n are quite large, the agreement between our solution and the data is rather good. Below $\sim 10$ GeV/n the error bars are much smaller and the scatter between data from different experiments is mainly due to solar modulation.
Notice that the spectrum has been corrected for the solar modulation using the widely used force-free approximation as in \cite{DiBernardo+2010}. During the ACE-CRIS data taking period, the solar wind potential $\Phi$ varied between 250 and 1000 MV, with an average value of $\Phi= 453$ MV \citep{Binns+2016}. We use this value for our calculation but, in order to quantify the impact of the solar modulation, in the same Figure~\ref{fig:Fe_spectrum} we report the Fe flux calculated with the maximum and minimum values of $\Phi$ during the relevant period: the variation is consistent with the observed scatter of the data.

\begin{figure}[t!!!!!!]
\begin{center}
\includegraphics[width=0.48\textwidth]{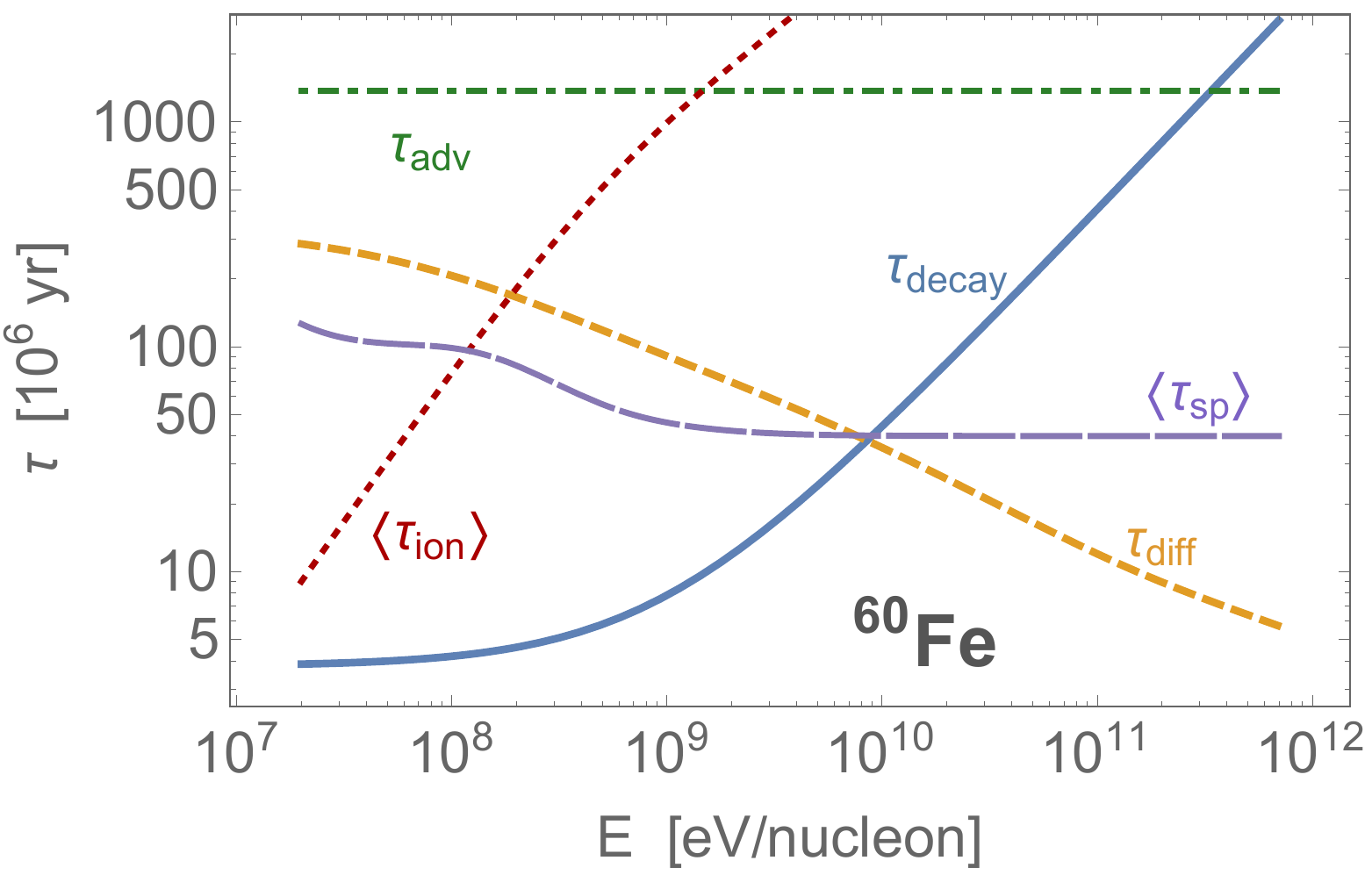}
\end{center}
\caption{Timescales in Myr for all processes involved in the transport of $^{60}$Fe {\bf for unmodulated energies}. The curves refer to the values of the model parameters reported in Table~\ref{table1}.}
\label{fig:timescales}
\end{figure}

Once the parameters of the model have been established, we can evaluate all the relevant timescales of the problem for both Fe isotopes. We plot the ones relevant for {\ferad\ } in Figure~\ref{fig:timescales}. Notice that the average timescales for spallation and ionization are calculated using the average target density, i.e. $\langle \tau \rangle = (\sigma v \, n_{\rm disc} h/H)^{-1}$. Figure~\ref{fig:timescales} makes it clear that, below $\sim 10$ GeV/n, the {\ferad } propagation is largely determined by radioactive decay. An analogous plot for {\festa} would show slightly different curves describing spallation, diffusion and ionization, due to the 7\% difference in atomic mass between the two isotopes. However, in the energy interval in which we are interested, diffusion is definitely more effective than advection, while spallation losses are more relevant than ionization above $\sim 100$ MeV/n. 

In deriving our solution for the particle propagation in  \S~\ref{sec:model} we neglected the effect of reacceleration. Such an assumption can be justified {\it a posteriori} estimating the reacceleration time as $\tau_{\rm reacc} = p^2/D_{pp}$ where the momentum diffusion is related to the spatial diffusion as $D_{pp}= p^2 v_A^2/(\eta_p D)$ and $\eta_p \simeq 0.1$ \citep{Drury-Strong2017}. In our model the reacceleration time  at 100 MeV/n is $\sim 4.5$ Gyr and increases $\propto R^{\delta}$ for larger rigidities, hence it is much larger than any other relevant timescale. This result reflects the low level of magnetic turbulence at scales that resonate with particles in this energy range, and it is in line with the idea that, in order to be effective, diffusive reacceleration scenarios generally require an uncomfortably large energy density in turbulence throughout the Galaxy \citep{Drury-Strong2017}.

%

\section{Results for the isotope ratio}
\label{sec:results}
\begin{figure}
\begin{center}
\includegraphics[width=0.48\textwidth]{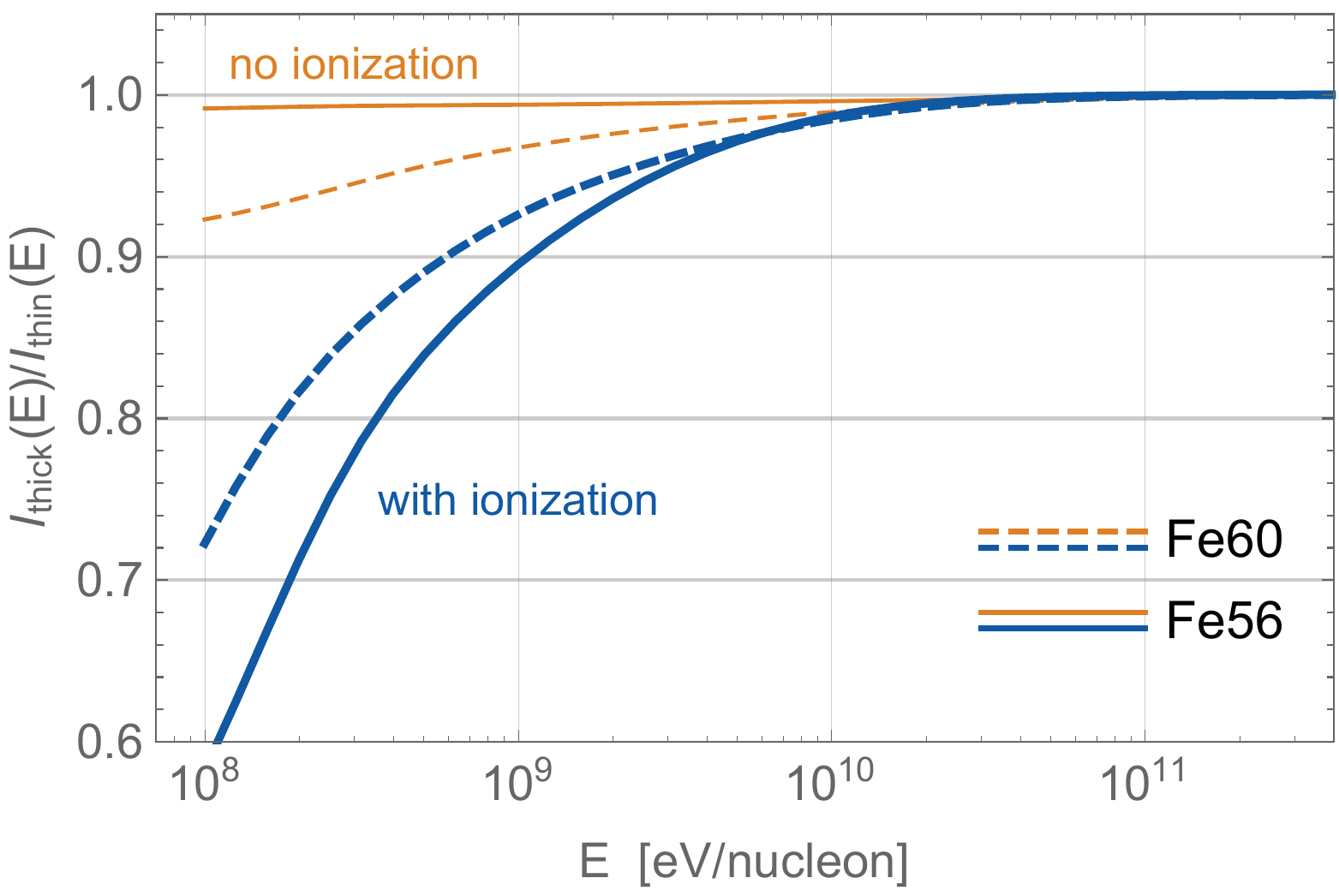}
\end{center}
\caption{Ratio between the Fe flux calculated in the thick disk model and that computed in the thin disk approximation. Both isotopes are shown: solid curves refer to \festa\ and dashed to \ferad, while thick (blue) curves include ionization losses and thin (orange) do not. The solar modulation is applied with a potential $\Phi= 453$ MV, as estimated by \cite{Binns+2016}.} 
\label{fig:Fe_thick_vs_thin}
\end{figure}

As already anticipated, the key parameter that allows us to understand the behaviour of the \ferad/\festa\ ratio is the different grammage experienced by the two isotopes. However, the grammage is a quantity that can be easily defined only in the thin disk model. Before discussing it, then, it is worth to consider the difference between the thick and the thin disk solutions. In Figure~\ref{fig:Fe_thick_vs_thin} we show the ratio between the fluxes calculated with the thick and the thin solutions, $I_{\rm thick}/I_{\rm thin}$,  for both \festa\ and \ferad. All curves are computed within the transport model described in the previous section. Notice that, for the thick disk solution, we are assuming that the diffusion coefficient in the disc is the same as the one in the halo. One can see that when the disk size is taken into account, the flux below $\sim 10$ GeV/n is suppressed by the fact that spallation reactions and ionization losses are more effective. The suppression reaches 30-40\% at $E \simeq 100$ MeV/n when all loss processes are included, while it is reduced to $\lesssim 10\%$  when ionization is not accounted for. This clearly highlights the importance of taking ionization losses into account, and also shows that at energies below $\sim 10$ GeV/n, the finite thickness of the disk cannot be neglected if one aims at computing CR fluxes with an accuracy better than few per cent. 

While the thin disk approximation leads to a non-negligible underestimate of absolute fluxes, it impacts the two Iron isotopes in a similar way, so that the error on the $I_{60}(E)/I_{56}(E)$ ratio is $\sim 15\%$ at $\sim 100$ MeV/n and decreases at larger energies. As a consequence, the thin disk approximation still provides a reasonably good estimate as far as the ratio of the two isotopes is concerned.

\begin{figure}[t!!!!!]
\begin{center}
\includegraphics[width=0.48\textwidth]{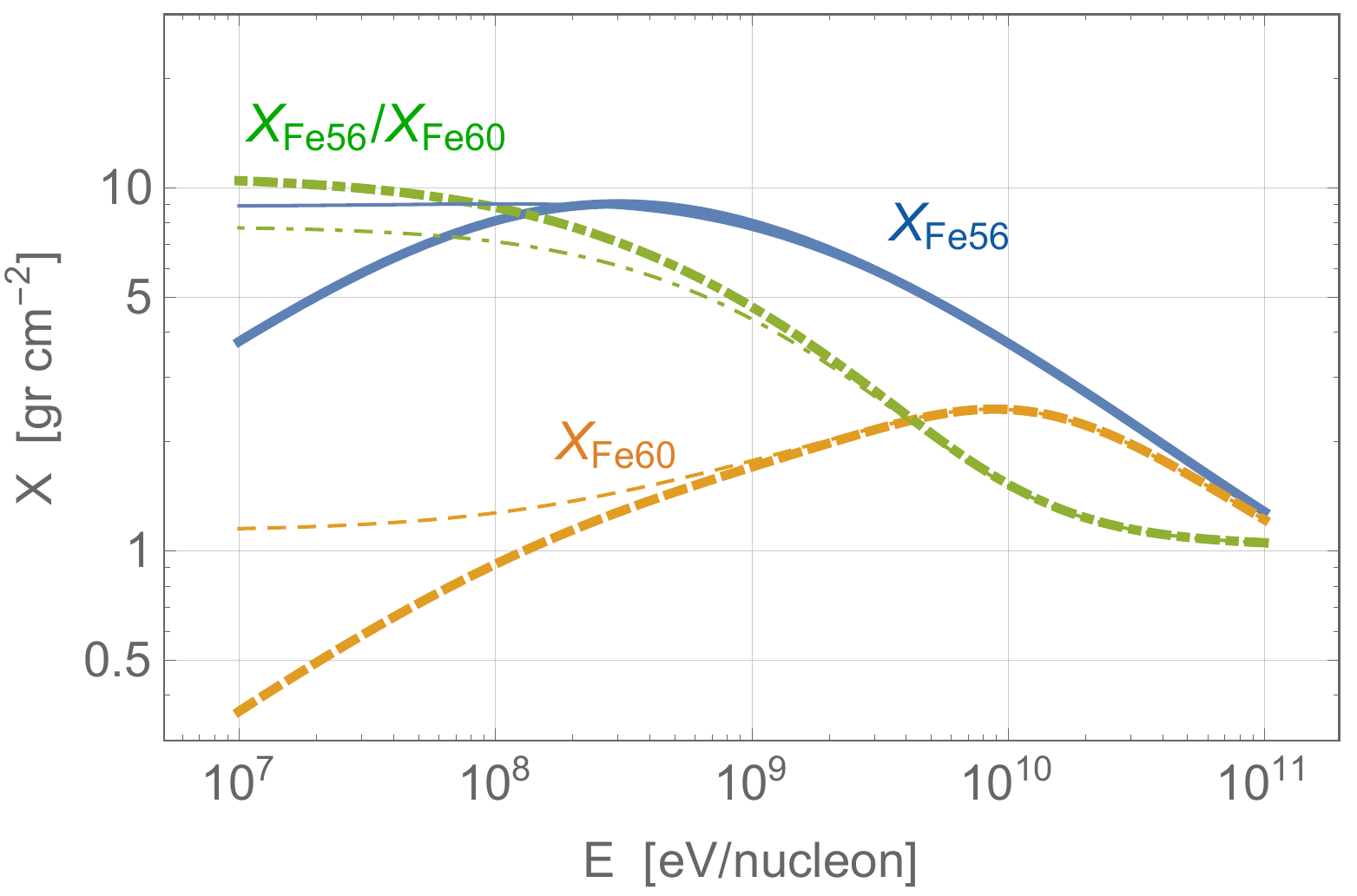}
\end{center}
\caption{Grammage accumulated by {\ferad\ } and {\festa\ } and ratio of the two grammages as a function of kinetic energy per nucleon. All lines assume the transport model described in \S\ref{sec:transport}. The solid (blue) curve is for $X_{\festa}$, the dashed (orange) curve for $X_{\ferad}$ and the dot-dashed (green) curve for $X_{\festa}/X_{\ferad}$. The latter quantity is clearly adimensional, but the numerical values on the y-axis still provide the right scale. Thin and thick lines refer to modulated and unmodulated energies, respectively.}
\label{fig:grammage}
\end{figure}

We then proceed to compute the grammage accumulated by \ferad\ and \festa\ (Eq.~\eqref{eq:X(E)}) within the thin disk approximation.  In Figure~\ref{fig:grammage}, we show $X_{\ferad}$, $X_{\festa}$, as well as the ratio of these two grammages. The plot shows the results for both unmodulated (thick lines) and modulated (thin lines) energies (with $\Phi= 453$ MV). 

It is clear that, at low energies, {\festa\ } suffers more spallation than {\ferad}. Only at energies $\gtrsim 10$ GeV/n the grammage accumulated by the two isotopes becomes equal, which correspond to the energy region where the decay time is larger than the diffusion time.
The results shown in Figure~\ref{fig:grammage} can be interpreted in terms of propagation lengths by means of Eqs.~\eqref{eq:xadv}-\eqref{eq:xdec}. As can be seen from Figure~\ref{fig:timescales}, at the low rigidities of the ACE-CRIS data, the decay time is much shorter than both $\tau_{\rm diff}$ and $\tau_{\rm adv}$. As a result, for {\ferad\ } Eq.~\eqref{eq:xdec} applies, while {\festa\ } falls in the case of Eq.~\eqref{eq:xdiff}, being $\tau_{\rm diff} < \tau_{\rm adv}$.
Therefore we expect
\begin{equation} \label{eq:X_ratio}
  \frac{ X_{\festa} }{X_{\ferad} } = \frac{H}{\sqrt{D \tau_d}} 
  	 \simeq 6.4 \,,
\end{equation}
which has been evaluated at the average (modulated) energy measured by CRIS, i.e. 327 MeV/n.
It is interesting to notice that the above ratio could reduce to unity even at low energies only in the advection dominated regime with very large advection speed. In fact in such a case Eq.~\eqref{eq:xadv} should be used for {\festa\ } and we would have $X_{\festa}/X_{\ferad}=\sqrt{D\tau_d}/(u_0\tau_d)$, which gives a result close to unity when diffusion and advection become of comparable importance. In our case $X_{\festa}/X_{\ferad}\approx 1$ at all energies requires $u_0 \gtrsim 500$ km s$^{-1}$.
This important fact implies that below such advection speed, at low energies transport is not fast enough to compete with decay: {\festa\ } nuclei live longer and always suffer more spallation and ionization losses than {\ferad}. The fluxes of the two are affected accordingly, with {\festa\ } undergoing stronger suppression.

\begin{figure}
\begin{center}
\includegraphics[width=0.48\textwidth]{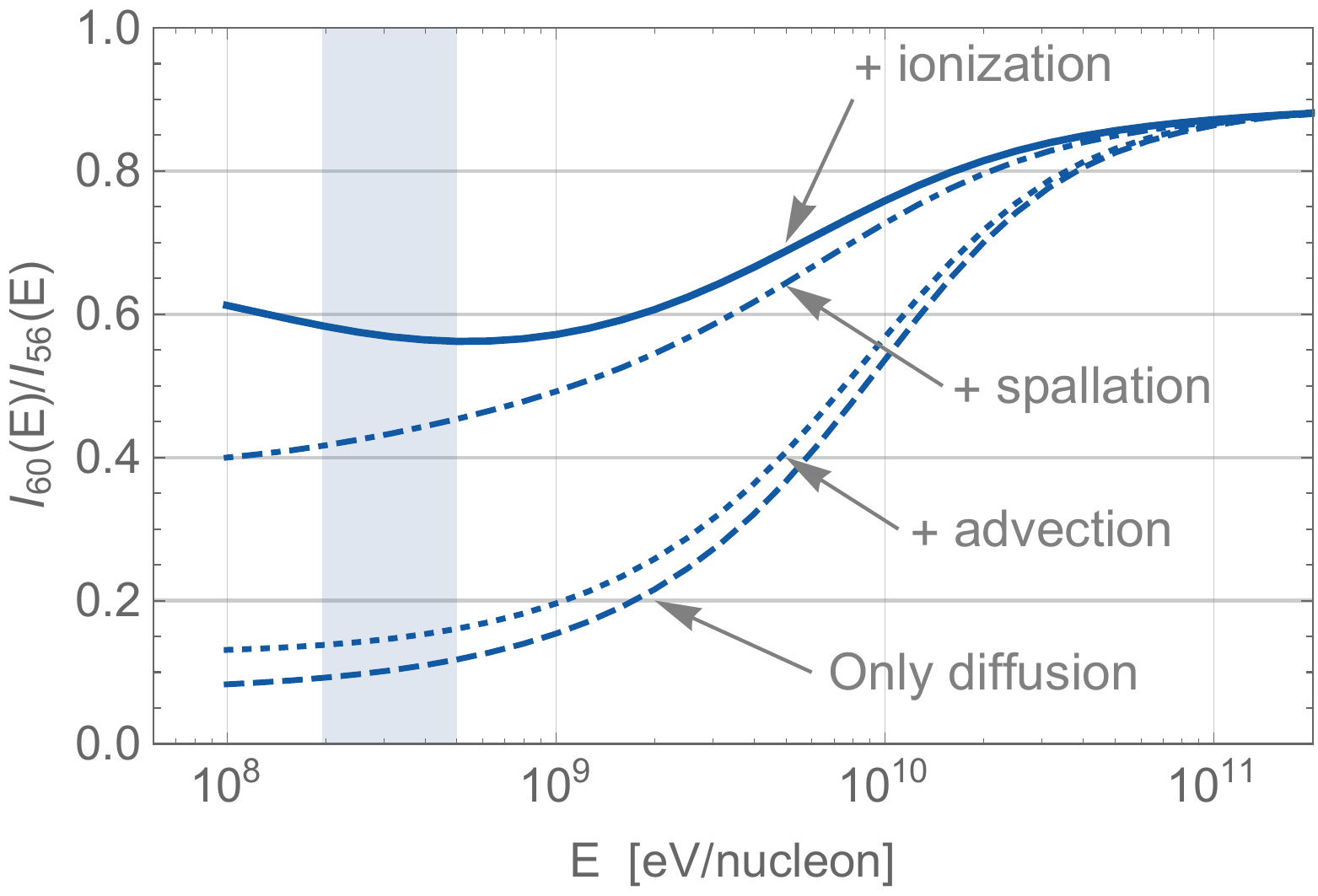}
\includegraphics[width=0.48\textwidth]{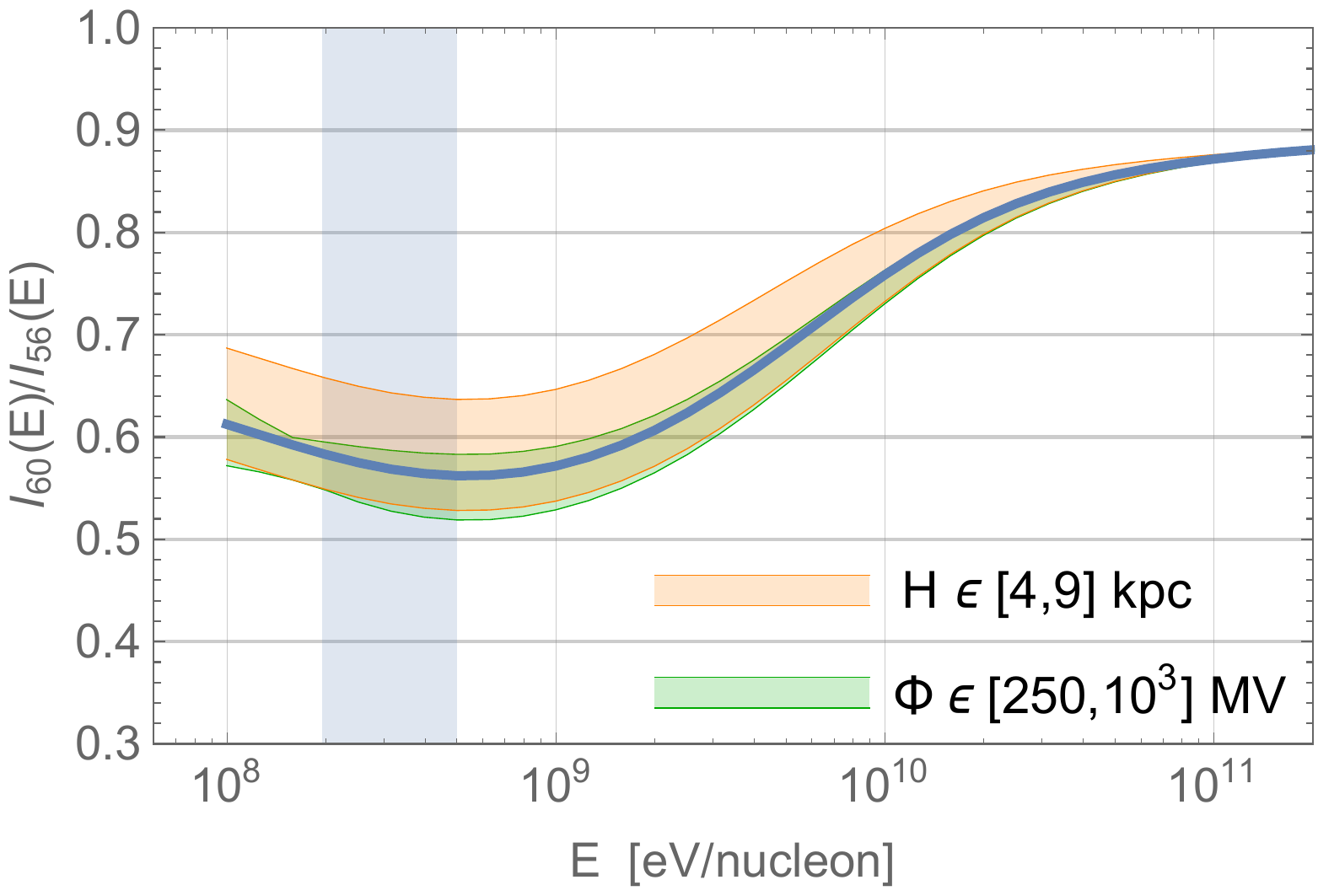}
\end{center}
\caption{Ratio between the fluxes of {\ferad\ } and {\festa\ } at the same energy per nucleon. \emph{Top panel}: Different lines show the role of each process during transport. From bottom to top, the different curves are computed accounting for: diffusion only (dashed), diffusion plus advection (dotted), spallation (dot-dashed) and ionization (solid -- full model).
The shaded vertical area shows the energy region of CRIS data. Solar modulation is taken into account with a potential $\Phi= 453$ MV.
\emph{Bottom panel:} As in the top panel, the solid line shows the flux ratio {\ferad\ / \festa\ } computed within our base-line model. The shaded bands represent how the result varies when the solar modulation changes between $\Phi= 250$ and 1000 MV (green band) and when the halo half thickness varies between 4 and 9 kpc (orange band). 
Notice that top and bottom panels have different vertical scales.}
\label{fig:ratio}
\end{figure}

What we just discussed helps us to understand the results showed in Figure~\ref{fig:ratio}, where we plot the propagated ratio between {\ferad\ } and {\festa}, namely $I_{60}(E)/I_{56}(E)$, under the assumption of an identical injection spectrum for the two ($q_{60}=q_{56}$). The solid curve shows the result obtained from Eq.~\eqref{eq:f0_sol_thick},  while the shadowed band shows the energy range of CRIS data. In order to illustrate the role of the different processes involved in propagation, in the same Figure we show the results that are obtained by including only part of the relevant processes: diffusion alone (dashed line - here the advection speed has been reduced by a factor 10), diffusion + advection (dotted line), diffusion + advection + spallation (dot-dashed line), diffusion + advection + spallation + ionization (solid line).
It is clear that if propagation were purely diffusive, the ratio between {\ferad\ } and {\festa\ } would be lower, the reason being that the escape time from the Galaxy would be longer and the {\ferad\ }would suffer more radioactive decays. Including advection decreases the residence time in the Galaxy and makes the fluxes of the two isotopes more similar. However, this is a minor correction in our model because the advection speed is only 5 km s$^{-1}$ (see Table~\ref{table1}). 
 On the other hand, when spallation and ionization losses are included, the {\ferad/\festa\ } ratio increases much more for the reason we discussed above: {\festa\ } experiences a larger grammage, hence suffering more losses than {\ferad}. As a consequence $I_{60}/I_{56}$ increases because $I_{56}$ is decreased. 
In conclusion the role of advection and much more the role of losses, cannot be neglected in this calculation.

After clarifying the role of the different processes, we now turn to the task of using CRIS measurement to deduce the ratio between {\ferad\ } and {\festa\ } in CR sources. CRIS measures the intensity ratio between {\ferad\ } and {\festa\ } at two slightly different energies $R_I=N_{60}(327 \, {\rm MeV/n})/N_{56}(340 \,{\rm MeV/n}) = (4.6 \pm 1.7) \times 10^{-5}$. We write the injection spectrum of CRs of species $s$ as $q_s(p)= n_0 \chi_s K_s p^{-\gamma}$, where $n_0$ is the gas density, $\chi_s$ is the relative abundance of each element and $K_s$ accounts for the efficiency of the injection process into the acceleration mechanism. In this notation, we define $R_{\rm source}= \chi_{60}/\chi_{56}$, so that the measured ratio $R_I$ between the two isotopes is connected to the  source ratio as
\begin{equation}
 R_I = R_{\rm source} \, \frac{K_{60}}{K_{56}} \, \frac{G_{60}(E_{60})}{G_{56}(E_{56})} \,.
 \label{eq:RQ}
\end{equation}
where $G_i$ accounts for propagation effects (i.e. $G_s(E)=I_s(E)/q_s(p)p^\gamma$).
The ratio $G_{60}/G_{56}$ is calculated using Eq.~\eqref{eq:f0_sol_thick} corrected for the Solar modulation and using $E_{60}= 327$ MeV/n and $E_{56}=340$ MeV/n.
Now, if one assumes that the injection efficiency is the same for both isotopes, namely that $K_{60} = K_{56}$, then the CRIS measurement translates into an abundance ratio $R_{\rm source} = (8.0 \pm 3.0) \times 10^{-5}$.

We notice, however, that the injection efficiency into the DSA mechanism may vary between different ions, being related to the mass to charge ratio \citep{Meyer-Drury-Ellison:1997,Ellison-Drury-Meyer:1997}. The matter is very far from settled and we will not discuss it in detail. We only notice that if one assumes, following the results from hybrid simulations by \cite{Caprioli+2017}, that injection efficiency is proportional to $\propto (A/Z)^2$, then {\ferad\ } is injected more efficiently than {\festa\ } by 15\%, so that the final result in terms of abundances is $R_{\rm source} = (6.9 \pm 2.6) \times 10^{-5}$.

Before concluding this section, it seems appropriate to discuss the impact on our results of two sources of uncertainties in our model: the size of the halo and solar modulation. We already mentioned that the halo size is estimated to be $\simeq 7$ kpc based on constraints from the Beryllium flux. Such a result can be affected not only by systematic errors in the Beryllium data, but also by uncertainties in the spallation cross sections \cite{Evoli[Be]+2020}. In the bottom panel of Figure~\ref{fig:ratio} we report how the estimated {\ferad\ / \festa\ } ratio changes varying the halo size between 4 and 9 kpc, while keeping the ratio $H/D_0$ constant. One can see that in the ACE-CRIS energy interval the uncertainty is $\sim 10\%$.
The same plot also shows the impact of changing the solar modulation, allowing the potential to vary between the maximum and minimum values experienced during the data acquisition by ACE-CRIS. In this case the uncertainty is less than 6\%. 

%

\section{Comparison with the {\it leaky-box} model}  \label{sec:leaky-box}

We think it is mandatory to compare the results presented in the previous section with those found in \cite{Binns+2016}, where the leaky-box model (LBM) was used to describe the transport. It is well known that such a model should be used with caution when dealing with unstable nuclei: the LBM is perfectly equivalent to the slab-diffusion model in describing stable particles, as showed by \cite{Ptuskin1974}, but it fails to describe unstable nuclei for the simple reason that particles can disappear from the system before reaching the boundary of the Galactic halo \citep{Ptuskin-Soutoul1998Rev}.

In spite of this important limitation the result presented in \cite{Binns+2016} is close to our finding within $\sim 8\%$. In the following we discuss the reasons for this.

In the LBM the general solution for the CR spectrum in the disk depends only on the characteristic timescales of escape, decay and spallation, and is written as
\begin{equation} \label{eq:LB}
  N = \frac{Q}{ \tau_{\rm esc}^{-1} + \tau_{\rm decay}^{-1} + \tau_{\rm sp}^{-1} } \,.
\end{equation}
Hence the Fe isotope ratio at the sources is 
\begin{equation} \label{eq:LB}
  Q_{60}/Q_{56} = (N_{60}/N_{56}) \times (\tau_{56} / \tau_{60})
\end{equation}
where $\tau_{56}^{-1}= \tau_{\rm esc}^{-1} + \tau_{\rm sp56}^{-1}$  and  $\tau_{60}^{-1}= \tau_{\rm esc}^{-1} + \tau_{\rm sp60}^{-1} + \tau_{\rm decay60}^{-1}$. 
The value of the escape time used by \cite{Binns+2016} was estimated, still in the framework of the LBM, based on the measurements of other radioactive nuclei \citep{Yanasak+2001} and is $\tau_{\rm esc} = 15 \pm 1.6$ Myr. Within the same model, the average gas density is $n_{\rm H+He}=0.38 \pm 0.04$, and this is used to calculate the spallation timescales leading to the values $\tau_{\rm sp56}= 4.45 \pm 0.47$ Myr and $\tau_{\rm sp60}= 4.27 \pm 0.45$ Myr. 

According to the analysis by \cite{Binns+2016}, CRIS measurements of \festa\ and \ferad\ refer to an average energy in interstellar space of 550 MeV/nu and 523 MeV/n, respectively. At those energies, our model gives $\langle \tau_{\rm sp56}\rangle = 56.1$ Myr and $\langle \tau_{\rm sp60} \rangle= 54.5$ Myr (see Figure~\ref{fig:timescales}).

The difference between the timescale estimated within our model and the LBM is a factor $\sim 12$, and is mainly due to the fact that our average density is $n_{\rm disk} h/H = 0.034$ cm$^{-3}$. Once the difference in average gas density is taken into account, the two estimates of the spallation timescales are still different by $\sim 12\%$, presumably due to differences in the adopted spallation cross section between the present work (see Appendix~\ref{sec:losses}) and that by \cite{Binns+2016}.

Aside from differences in the spallation and escape time-scales, another difference between this work and that by \cite{Binns+2016} is that the latter neglects ionization losses, which in our calculation turn out to be non-negligible, being the ionization loss time of the same order of $\tau_{\rm sp}$  for energies $\sim 100$ MeV/n, which translates into a correction of $\sim 20\%$ to the final \ferad/\festa\ ratio in the CRIS energy band as shown in Figure~\ref{fig:ratio}.

Using our estimated time-scales in the leaky box expression connecting the measured and source ratio between the isotopes (Eq.~\ref{eq:LB}), one would estimate $R_{\rm source, LBM}\approx 4\times 10^{-4}$, a factor of 5 larger than the estimate by \cite{Binns+2016}.

In summary, while our estimate and the estimate by \cite{Binns+2016} of the $\ferad/\festa$ ratio at the sources turn out to be very close, they correspond to very different physical conditions for CR propagation. In particular, in our model CR particles have much longer residence times in the Galaxy and lose a non-negligible fraction of their energy while propagating through the low-density halo. The similarity between the two estimates of $R_{\rm source}$ seems at present only a puzzling coincidence. What actually enters the relation between the source ratio and the measured ratio between isotopes is a {\it survival probability} \citep{Lipari2014}. This is coincident with a ratio between timescales, as adopted in the LBM, only when the confinement volume is coincident with the volume occupied by the sources. When the former is much larger than the latter, instead, one finds in general, that a fraction of the actual escape time (determined by the ratio between source and confinement volume \cite{Lipari2014}) enters into Eq.~\eqref{eq:LB}. 

In fact, our estimate of the diffusion time is the same that allows one to reproduce the total Beryllium flux \cite{Evoli[Be]+2020}. Our conclusion, in agreement with past works \cite[e.g.][]{Jones+2001}, is that the confinement times estimated by \cite{Yanasak+2001} are all underestimated by a factor of order 10.

\section{Summary and Conclusions}	\label{sec:conc}
In this work we modelled the propagation of Iron nuclei through the Galaxy within the disk/halo diffusion model in order to translate the \ferad/\festa\ ratio measured by ACE-CRIS in CRs into an estimate of the relative abundance of the two isotopes in CR sources.  Following \citep{Evoli[Be]+2020}, the parameters of the transport model have been fixed in such a way as to reproduce the fluxes of CR $p$, He, C, N, O  plus the ratios Be/C, B/C, Be/O and B/O as measured by AMS-02. In addition, we adopted a halo size of $\sim 7$ kpc as estimated from the CR Beryllium flux \citep{Evoli[Be]+2020}. 

At energies $\lesssim 1$ GeV/n, where the ACE-CRIS measurements have been performed, the CR transport is determined by several processes: diffusion, advection, spallation, ionization losses and solar modulation. We accounted for all these processes, quantifying the role of each one in determining the {\ferad/\festa\ } ratio. 

We showed that  at energies $\lesssim 1$ GeV/n also the size of the Galactic disk becomes important, being comparable with the energy loss length of heavy nuclei. Hence, we have explicitly accounted for the disk size in our analytical description of the CR transport showing that, under the assumption that the diffusion coefficient is the same as in the Galactic halo, the Fe flux is suppressed by $\sim 30\%$ with respect to the infinitely thin disk approximation. On the other hand, the final {\ferad/\festa\ } is affected only by $\sim 10\%$ because the two isotopes are affected in a similar way.
Finally, we also accounted for the preferential injection of heavier nuclei in the shock acceleration mechanism.

Within the above scenario we found for the {\ferad/\festa\ } ratio at the CR sources \ferad/\festa\ = $(6 \div 11) \times 10^{-5}$ (accounting for both measurement errors and model uncertainties).
Such a value is especially interesting when compared with the average abundance in the ISM, which is $\sim 3 \times 10^{-7}$, implying that the CRs detected at Earth cannot be produced by accelerating only the average ISM composition. As a consequence, and not surprisingly, we can exclude the blast waves of type Ia SNe as the main source of Galactic CRs, in that they mainly accelerate material from the average ISM. Our result requires, instead, that some fraction of the accelerated material should come from fresh SN ejecta (where {\it fresh} means much younger than the \ferad\ decay time).  
The exact amount of accelerated fresh ejecta is non-trivial to estimate, because the \ferad\ yield from SN explosions depends on the progenitor initial mass \citep{Woosley-Weaver1995} as well as the star rotational speed \citep{Chieffi-Limongi2013}. The value of the ratio \ferad/\festa\ ranges between $2\times10^{-4}$ and $8\times10^{-3}$ \citep{Woosley-Weaver1995}, hence one can infer that the amount of fresh ejecta that needs to be accelerated should be a fraction between few percent and few tens of percent of the total accelerated material.

The two main scenarios in which this can be realised are one in which acceleration occurs at the reverse shock of the SN explosion and one in which the fresh ejecta of an explosion are accelerated by the forward shock of a second nearby event. A possible way to disentangle between these two possibilities is by looking at the abundances of other nuclei, especially the $^{22}$Ne, whose over-abundance with respect to the Solar one is still not completely understood \cite[see][for a critical discussion]{Prantzos2012}.

A major surprise is that our results are in agreement with the estimate obtained by \cite{Binns+2016}. The latter work adopted a leaky box description of particle transport, which is in principle not appropriate to describe the propagation of unstable nuclei, and in addition neglected advection and ionization losses, while we find the latter to be very relevant. Our scenario predicts a confinement time $\sim 10$ times larger than  the LBM. As we discussed in \S~\ref{sec:conc}, this is likely the key to understand the incidental agreement. When the volume of the sources is only a fraction of the total confinement volume, a LBM description of the survival probability becomes appropriate for a confinement time which is a fraction of the actual one.  
We have shown that using the correct escape time, the LBM provides a result for the Fe isotope ratio at the sources $\sim 5$ times larger than our estimate.

It is worth stressing that while in the present study we use the propagation model to constrain the {\ferad\ } abundance at the sources, analogous measurements performed for other radioactive secondaries produced during the spallation process, like $^{10}$Be or $^{14}$C, could be used in the opposite direction, namely to provide a valuable test of the CR propagation regime at low energies.

\begin{acknowledgments}
We acknowledge the International Space Science Institute (ISSI) for Teamwork 351 on ``{\it The origin and composition of galactic cosmic rays}'', where this work was conceived. We are especially indebted to N. Prantzos, B. Binns, V. Ptuskin and V. Tatischeff for useful discussion and comments. We also acknowledge the hospitality of the Kavli Institute in Santa Barbara, where part of the work was carried out. This research was supported by Grants ASI/INAF n. 2017-14-H.O,  SKA-CTA-INAF 2016, INAF {\it Mainstream} and by the National Science Foundation under Grant No. NSF PHY-1748958.
\end{acknowledgments}

\appendix
\section{energy losses} \label{sec:losses}

For the total spallation cross section we use the following expression from \cite{Letaw+1984}:
\begin{eqnarray} \label{eq:sigma_sp}
  \sigma_{\rm sp}(E_k) = 45 \, A^{0.7}  \, 10^{-27} \rm cm^2                            \hspace{4.2cm}		\\ \nonumber 
  	\times  \left[1 + 0.016 \sin\left(5.3 - 2.63 \ln(A)\right) \right]    \hspace{3cm}		\\ \nonumber
	\times  \left\{1 - 0.62 \exp\left(- \frac{E_k}{2 \cdot 10^8} \right) \sin\left[10.9 \left(\frac{E_k}{10^6} \right)^{-0.28} \right] \right\}
   \,,
\end{eqnarray} 
where $E_k$ is the kinetic energy per nucleon and $A$ is the bullet's atomic masses (the target is assumed to be purely protons). According to \cite{Letaw+1984}, the mean error of Eq.\eqref{eq:sigma_sp} is less than 5\% for energies above 100 MeV/n.
Notice that the spallation cross section use in \citep{Binns+2016} is the one measured by \cite{Westfall+1979}: we notice that at the energy where such measurements were performed, namely 1.88 GeV/nucleon, their result for the total inelastic cross section of Fe onto H target is $\sim 9\%$ smaller than the value given by Eq.~\eqref{eq:sigma_sp}. The extrapolation at lower energies could be responsible for the $\sim 20\%$ difference in the calculated spallation timescale as discussed in \S~\ref{sec:leaky-box}.

For the energy losses due to ionization we use an interpolation formula provided by \cite{Mannheim-Schlickeiser1994} (see their equations [4.32]-[4.34]), which is proportional to the energy losses of protons and is valid when the energy per nucleon is $E_k \lesssim 1$ TeV:
\begin{equation} \label{eq:dEdtA_ion}
  \left( \frac{dE}{dt} \right)_{\rm ion, Z}  = Z_{\rm eff}^2 \left( \frac{dE}{dt} \right)_{\rm ion, p}
\end{equation} 
where the effective charge of the nucleus is given by $Z_{\rm eff}= Z (1- 1.034 \exp[-137 \beta Z^{-0.688}] )$ and the energy losses of protons are:
\begin{eqnarray} \label{eq:dEdtH_ion}
  \left( \frac{dE}{dt} \right)_{\rm ion, p} = 
  	1.82 \cdot 10^{-7}  \, \left( [n_{\rm HI} + n_{\rm H_2}]/{\rm cm^{-3}} \right)	\nonumber	\\
	\times \left[ 1 + 0.0185 \ln(\beta) \Theta(\beta-\beta_0) \right]
  	\frac{2 \beta^2}{\beta_0^3 + 2\beta^3}  \; {\rm eV \, s^{-1}} \,.
\end{eqnarray} 
$\beta_0 = 0.01$ is the minimum Lorentz factor such that Eq.~\eqref{eq:dEdtH_ion} is valid.
The momentum loss function used in Eq.~\eqref{eq:transport0} is $\dot{p}_{\rm ion} = dp/dt = dp/dE \times (dE/dt)_{\rm ion}= A/v \times (dE/dt)_{\rm ion}$.

\bibliography{Iron60}

\end{document}